\documentclass[aip,jcp,reprint]{revtex4-1}
\usepackage{amsmath}
\usepackage{amsfonts}
\usepackage{amssymb}
\usepackage{mathrsfs}
\usepackage{txfonts}
\usepackage{braket}
\usepackage[version=3]{mhchem}
\usepackage{graphicx}
\usepackage{color}
\usepackage{upgreek}
\usepackage{bm}
\usepackage{multirow}
\usepackage{natbib}
\usepackage{here}
\usepackage{tikz}
\usepackage{physics2}
\usepackage{float}
\allowdisplaybreaks
\usepackage[top=25truemm,bottom=20truemm,left=20truemm,right=20truemm]{geometry}
\usepackage{hyperref}
\usepackage[legacycolonsymbols]{mathtools}
\usepackage{xr}
\begin{document}

\title{
  On transition path times for condensed-phase non-adiabatic electron transfer reactions under a two-parabola model
}

\author{Ryo Nihei}
\affiliation{Department of Chemical Science and Technology, Graduate School of Engineering, Kyoto University, Kyoto, 615-0510, Japan}

\author{Hiroki Uratani}
\email{uratani@moleng.kyoto-u.ac.jp}
\affiliation{Department of Chemical Science and Technology, Graduate School of Engineering, Kyoto University, Kyoto, 615-0510, Japan}
\affiliation{PRESTO, Japan Science and Technology Agency, Kawaguchi, Saitama, 332-0012, Japan}

\author{Hirofumi Sato}
\affiliation{Department of Chemical Science and Technology, Graduate School of Engineering, Kyoto University, Kyoto, 615-0510, Japan}
\affiliation{Fukui Institute for Fundamental Chemistry, Kyoto University, Kyoto, 606-8103, Japan}

\begin{abstract}
Condensed-phase non-adiabatic electron transfer (ET) reactions play an important role in various areas of science. 
The characteristics of the microscopic transition processes involved in these reactions and their dynamic behavior remain much room for further understanding.
In this study, we performed non-adiabatic dynamics simulations based on Zhu-Nakamura theory, combining the two-parabola model with Kramers-like Langevin dynamics.
We numerically analyzed the dependence of the macroscopic reaction time and the timescale of the microscopic transition process on the diabatic coupling between the diabatic states and on the friction parameter in Langevin dynamics.
Focusing on resulting two quantities, that is, the reaction rate and the transition path times (TPTs), which characterize the macroscopic population transfer and the individual transition process, respectively, we discuss how non-adiabatic transitions influence the reaction dynamics.
\end{abstract}

\maketitle

\section{INTRODUCTION}

Condensed-phase non-adiabatic electron transfer (ET) reactions play a key role in a wide range of chemical, biological, and materials systems, including photochemical charge separation, redox catalysis, and bimolecular electron exchange. 
A microscopic understanding of these reactions is important for new technologies such as photoredox catalysis, solar energy conversion, and artificial photosynthesis.\cite{Li2024, Pea2025, Xia2026, Weng2020, Chen2025, Cai2025, Shi2025, Benk2022, Hassaan2023} 
Although the mechanisms and dynamics of condensed-phase ET reactions have been widely investigated from both theoretical and experimental perspectives, the coupling among nuclear motion, solvent fluctuations, and electronic transitions continues to pose fundamental problems.\cite{Kim1990, Kim1992, Ando1991, Blumberger2015, Cohen2003, Liu2025, Wu2025}
In particular, the characteristics of the microscopic transition processes involved in these reactions and their dynamic behavior remain poorly understood. 

To describe the non-adiabatic electron transer in condensed phases, one has to properly take into account thermal fluctuation and transitions between different electronic states. In the present study, we performed non-adiabatic dynamics simulations\cite{Tully1971, Tully1990, Barbatti2022, Prezhdo1997, Landry2012} based on the combined use of Kramers-like Langevin dynamics\cite{Kramers1940, Blumberger2015, Nitzan2006Book} and Zhu-Nakamura theory\cite{Zhu1992, Zhu1992_, Zhu1993, Zhu1994, Zhu2001}-based treatment of non-adiabatic transitions to study the microscopic dynamics of condensed-phase non-adiabatic ET reactions. The model system is described by a two-parabola model.
Note that Zhu-Nakamura theory provides an appropriate non-adiabatic transition probability even for curve-crossing problems of two potentials with opposite signs of slopes, where the Landau-Zener formula becomes less satisfactory.\cite{Zhu1992}
By combining Langevin nuclear dynamics with explicit electronic transitions, our approach provides a unified microscopic description of coupled nuclear-electronic motion in condensed-phase ET.

To characterize the microscopic dynamics, we focus on the transition path time (TPT), which quantifies the timescale of such microscopic transition processes.\cite{Laleman2017, Medina2018, Satija2017, Sharma2022}
TPT represents the time required for a trajectory to cross the energy-barrier region along the reaction coordinate and provides microscopic information that differs from macroscopic reaction rates.
So far, research on TPT has been conducted mainly in the adiabatic regime, where analytical distributions have been obtained by approximating the energy barrier as an inverted parabola.\cite{Laleman2017} 
The TPT has been extensively studied particularly in the field of biomolecular conformational changes.\cite{Laleman2017, Shivangi2022}
Here, we employ the TPT concept to characterize electron-transfer reactions.
We analyze the effects of the diabatic coupling $V$ and the friction parameter $\gamma$ on the TPT distribution and the average TPT, and examine the relationship between these microscopic time scales and the macroscopic reaction time $(\tau)$.

\section{METHODS}

\subsection{Model system\cite{Landry2012}} \label{subsec_model}
\begin{figure}
    \centering
    \includegraphics[width=3in]{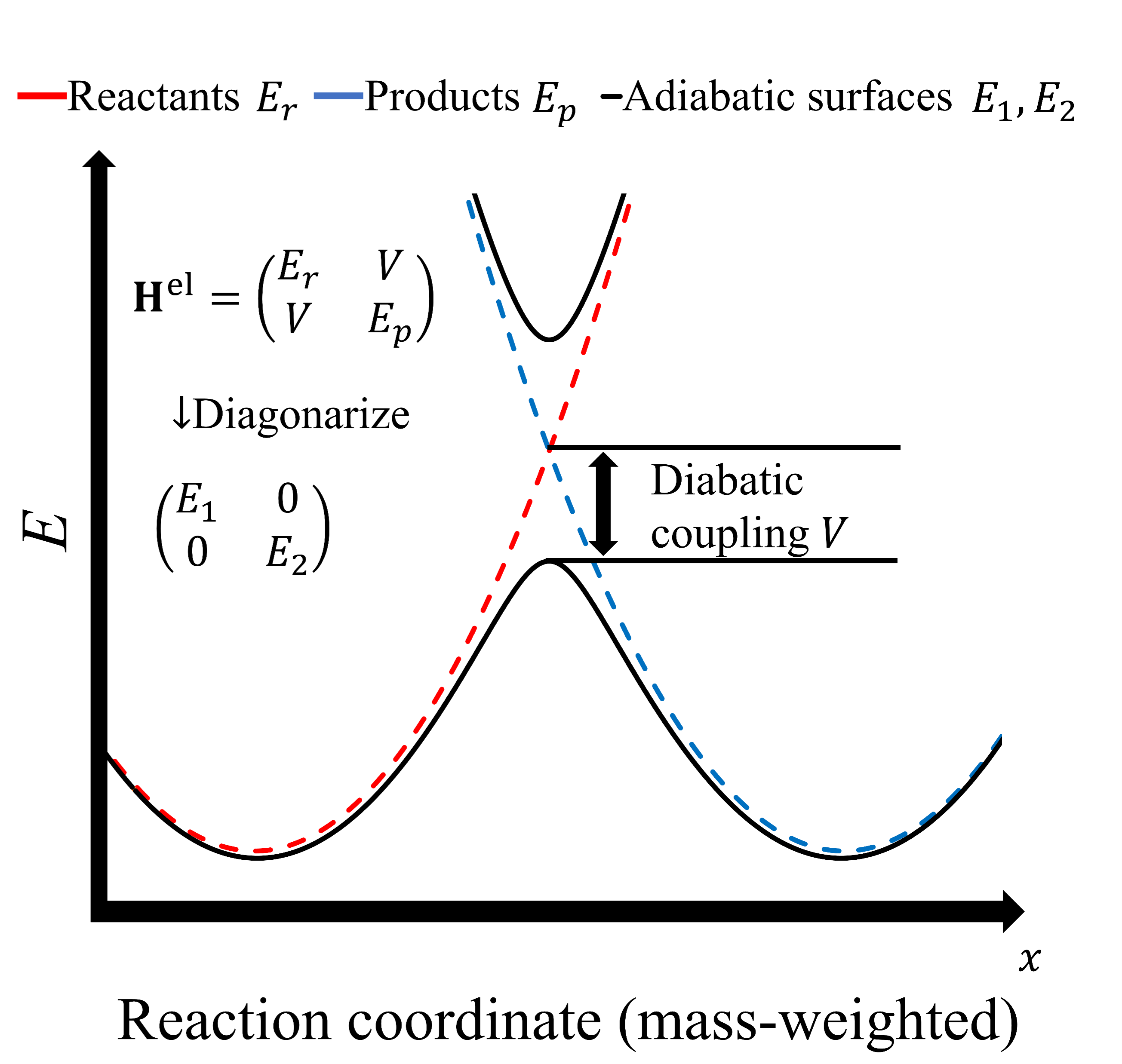}
    \caption{Two-level avoided crossing model. The red and blue dashed lines represent reactants and products with the diabatic coupling $V$, whereas the black solid lines are the adiabatic states.
    The reaction proceeds from left to right well.}
    \label{fig:2-level}
\end{figure}
We describe the condensed-phase electron transfer reaction by a two-parabola model. 
As a representation of a two-level system coupled to a bath, the two-parabola model has played a key role in theoretical analyses of electron-transfer processes.\cite{Hazra2010, Mller1997, Legget1987, Makarov1993, mac2002, BenNun2007, Landry2012}
Two diabatic parabolic potential energy surfaces represent the reactants $(E_r)$ and products $(E_p)$, with the diabatic coupling $V$ between these states(Fig.\ref{fig:2-level}).  
The electronic Hamiltonian is
\begin{align}
    &\bold{H}^{\rm{el}} =\begin{pmatrix}
        E_r && V \\
        V && E_p
    \end{pmatrix}= \begin{pmatrix}
        \frac{1}{2}\omega^2(x+x_\mathrm{a})^2 && V \\
        V && \frac{1}{2}\omega^2(x-x_\mathrm{a})^2
    \end{pmatrix}, \label{hamiltonian}
\end{align}
where $\omega$ is the harmonic frequency, $x$ is the mass-weighted reaction coordinate, and $\mp x_\mathrm{a}$ is the most stable point of the reactants and products.
To simplify the discussion of TPTs (see \ref{subsec_tpt}), we set the two diabatic curves to be symmetric, i,e., the reaction driving force is 0.
To check the dependence on the driving force, we also performed the simulations with driving forces $\Delta G=\pm 10\,{\rm meV}$, where $\Delta G$ stands for the bottom of the product well minus the bottom of the reactant well, and no considerable differences were seen in the results (see Section S6 in Supplementary Informatio	n).
The adiabatic energies, which are given by diagonalizing the Eq.\eqref{hamiltonian}, are 
\begin{align}
    E_1(x) &= \frac{1}{2}\left(E_r(x)+E_p(x)\right) - \frac{1}{2}\sqrt{\left(E_r(x)-E_p(x)\right)^2+4V^2} \\
    E_2(x) &=  \frac{1}{2}\left(E_r(x)+E_p(x)\right) + \frac{1}{2}\sqrt{\left(E_r(x)-E_p(x)\right)^2+4V^2}.
\end{align}
The adiabatic basis $\ket{\Phi_1}$ and $\ket{\Phi_2}$ are the eigenvectors of the Hamiltonian Eq.\eqref{hamiltonian} and can be expressed as a linear combination of the diabatic basis $\ket{\Xi_r}$ and $\ket{\Xi_p}$
\begin{align}
    \ket{\Phi_1} &= c_{r,1}(x)\ket{\Xi_r} + c_{p,1}(x)\ket{\Xi_p} \\
    \ket{\Phi_2} &= c_{r,2}(x)\ket{\Xi_r} + c_{p,2}(x)\ket{\Xi_p},
\end{align}

where $c_{r,1}(x)$, $c_{p,1}(x)$, $c_{r,2}(x)$, and $c_{p,2}(x)$ are
\begin{align}
    c_{r,1}(x) &= \sqrt{\frac{1}{2}-\frac{1}{2}\frac{1}{\sqrt{1+\eta(x)^2}}}, \ c_{p,1}(x) = -\sqrt{\frac{1}{2}+\frac{1}{2}\frac{1}{\sqrt{1+\eta(x)^2}}} \notag \\
    c_{r,2}(x) &= \sqrt{\frac{1}{2}+\frac{1}{2}\frac{1}{\sqrt{1+\eta(x)^2}}}, \ c_{p,2}(x) = \sqrt{\frac{1}{2}-\frac{1}{2}\frac{1}{\sqrt{1+\eta(x)^2}}} \notag \\
    \eta(x) &=\frac{2V}{E_r(x)-E_p(x)}.\label{combination}
\end{align}

 We will use these coefficients in \ref{subsec_caharacterization}.

\subsection{Dynamics\cite{Laleman2017, Landry2012}} \label{subsec_dynamics}

To simulate the dynamics under the solvent fluctuation, we use the Langevin equation of motion 
\begin{align}
    \ddot{x}(t) = -\frac{dE_i(x)}{dx} - \gamma \dot{x}(t) + \xi(t),\label{7}
\end{align}
 where $i$ is the index of the active surface (1 or 2), $\gamma$ is a friction parameter and $\xi$ is a Gaussian random force. $\xi$ satisfies the fluctuation--dissipation relation 
 \begin{align}
     \left\{
     \begin{matrix}
         \langle \xi(t) \rangle =  0 \\
         \langle \xi(t)\xi(t') \rangle = 2\gamma k_\mathrm{B}T \delta(t-t'),
     \end{matrix}
     \right.\label{8}
 \end{align}
 where $\langle...\rangle$ denotes the ensemble average. 
 $-\gamma \dot{x}$ represents the energy dissipation, and $\xi(t)$ represents thermal fluctuations.
 As shown in Eq.\eqref{8}, the standard deviation of $\xi$ is proportional to $\sqrt{\gamma}$.
 The larger $\gamma$ corresponds to the diffusive regime, while the smaller to the ballistic regime.

 To account for non-adiabatic transitions, we used Zhu-Nakamura theory,\cite{Zhu1992, Zhu1992_, Zhu1993, Zhu1994, Zhu2001} which provides an exact solution for transitions in a one-dimensional two-level system.
Furthermore, in the specific context of the present work, the advantage of the use of Zhu-Nakamura theory is that it simplifies the analyses since it represents the non-adiabatic effects with transitions at the diabatic potetial crossing point, whose probability solely depends on the incident momentum. In contrast, in Tully's fewest-switches surface hopping (FSSH)\cite{Tully1990}, non-adiabatic transitions can occur anywhere in the coordinate space, and the transition probability depends on the time-dependent electronic wavefunction carrying history dependence, adding extra complexity to the characterization of trajectories.
Note that we also performed the FSSH-based simulations with and without including the decoherence correction\cite{Zhu2004, Zhu2005, Zhu2004_, Granucci2007, Cheng2008}, where the results of the latter are qualitatively consistent with those obtained using Zhu-Nakamura theory (see Section S4 in Supplementary Information).

 In Zhu-Nakamura theory, the non-adiabatic hopping probability $P_\mathrm{hop}$ is given by
\begin{align}
    P_{\rm{hop}} = \exp\left({-\frac{\pi}{4\sqrt{a^2}}\sqrt{\frac{2}{b^2+\sqrt{b^4-1}}}}\right). \label{9}
\end{align}
In this study, because diabatic gradients have opposite signs at the avoided crossing point, the sign after $b^4$ is minus.
The parameters $a$ and $b$ represent the effective coupling and the effective collision energy, respectively, where
\begin{align}
    a^2 = \frac{\hbar^2}{2}\frac{\sqrt{|F_2F_1||F_2-F_1|}}{(2V)^3},\label{10} \\
    b^2 = (E_t-E_X)\frac{|F_2-F_1|}{\sqrt{F_2F_1}(2V)},\label{11}
\end{align}
where $F_1$ and $F_2$ are the diabatic gradients, $E_t$ is the total energy and $E_X$ is the diabatic potential energy at the avoided crossing point.
As shown in Eq.\eqref{9}-\eqref{11}, the larger the momentum (higher total energy) at the avoided crossing point is, the higher the hopping probability is.
Moreover, as the diabatic coupling $V$ increases, the hopping probability decreases $(P_{\rm{hop}}\varpropto\exp{(-V^2)})$.
Thus, under a large value of $V$, a trajectory passing through the avoided crossing region is likely to entail a {\it diabatic} transition from the reactant diabatic state to the product diabatic state.

In this study, we started the simulation at equilibrium in the reactant state (left parabola). 
In other words, we sampled the initial coordinates from the Boltzmann distribution in $E_r$, and the initial momenta were sampled from the Maxwell-Boltzmann distribution.
The probability of choosing the bottom adiabatic surface as the active surface is 
\begin{align}
    |c_{r,1}(x)|^2
\end{align}

\subsection{Reaction time constant ($\tau$) and transition path time (TPT)} \label{subsec_caharacterization}
As stated in \ref{subsec_model}, the adiabatic states consist of linear combinations of diabatic states. Following Ref.\citenum{Zhu2001}, we define $|c_{r,i}|^2$ as the probability of being the reactant state.
If the active surface is $E_i$ corresponding to $\ket{\Phi_i(x)}$, the probability of being in the reactant state is $|c_{r,i}(x)|^2$. 
The population of reactants $[\mathrm{R}]$ is given by averaging this probability over all trajectories at each time step.
Since the potential energy surfaces are symmetric with respect to the reactant--product permutation, the reaction scheme of this system is
\begin{align}
    \rm{Reactants} \overset{\text{\it{k}}}{\underset{\text{\it{k}}}{\rightleftharpoons}} Products,
\end{align}
where $k$ is a rate constant.
Thus, the time evolution of $[\mathrm{R}]$ can be written as
\begin{align}
    [\mathrm{R}](t) =\langle |c_{r, \alpha}(x)|^2 \rangle(t)= \frac{1}{2}\left(1+\exp(-2kt)\right), \label{14}
\end{align}
where $\alpha$ is active state index and $t$ is time.
Assuming that each reactant $\rightleftharpoons$ product transition event occurs independently of the others, the reaction can be modeled as a Markov process. 
In this case, the time constant $\tau \ (= 1/k)$ corresponds to the mean first-passage time (MFPT), defined as the mean time required for a system in thermal equilibrium on the reactant side to reach the product state for the first time.\cite{Hanggi1990}


\begin{figure}
    \centering
    \includegraphics[width=3in]{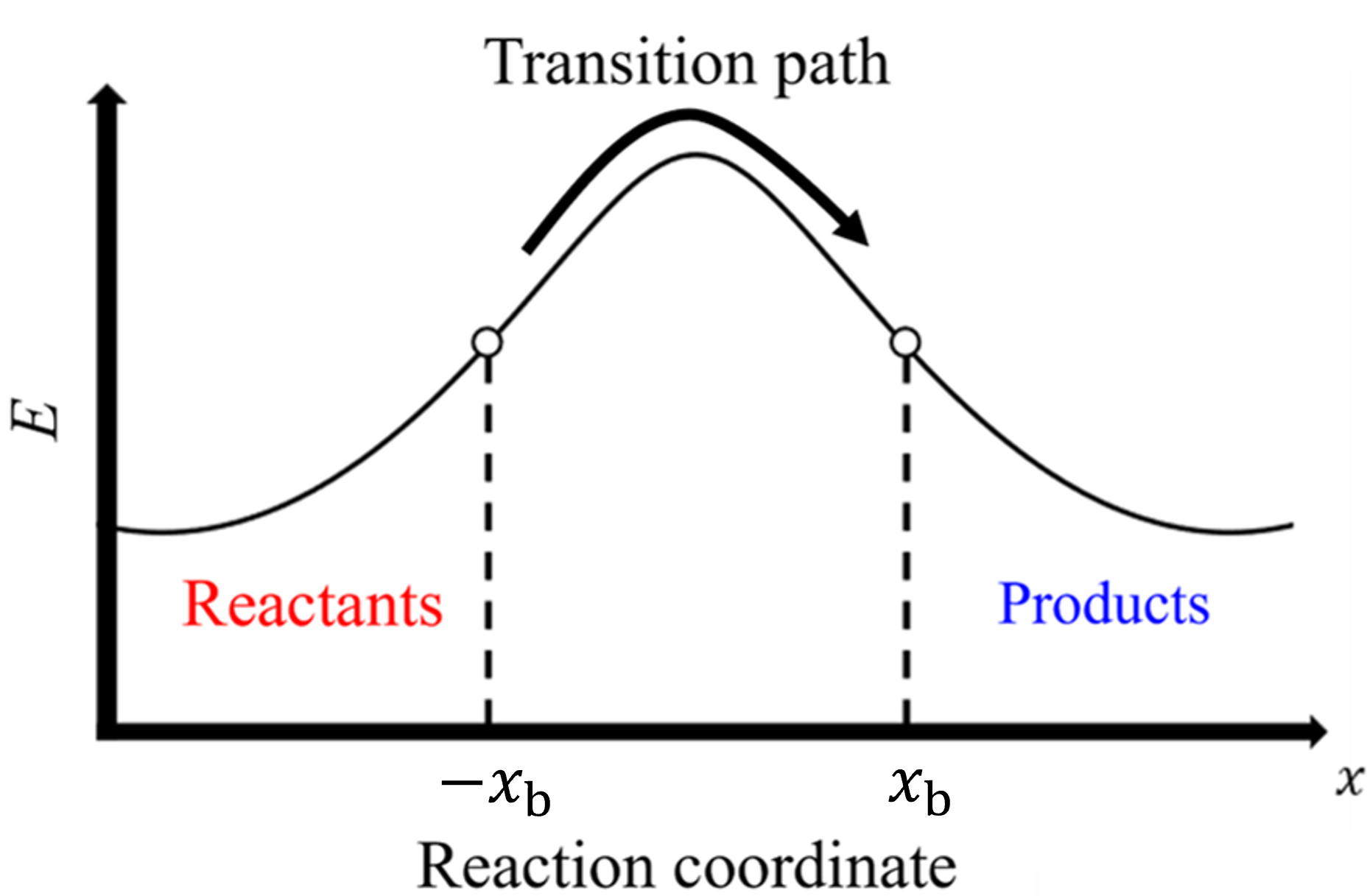}
    \caption{Schematic diagram of a transition path (the transition path represents a microscopic transition process)}
    \label{fig:placeholder}
\end{figure}

In contrast, the transition path time (TPT) characterizes the timescale of microscopic transition processes in which an individual trajectory passes through an energy-barrier region, in contrast to the timescale of population transfer characterized by $\tau$.
In the following, we explain how the TPTs were calculated from the simulated trajectories.
The model energy surfaces are symmetric with respect to the interchange of reactants and products. 
Although an adiabatic state is a linear combination of the reactant and product diabatic states (Eq.\eqref{combination}), the lower adiabatic state is predominantly composed of the reactant diabatic state at $x=-x_\mathrm{b}$ and the product diabatic state at $x=x_\mathrm{b}$.
Hence, we defined the TPT as the time required by the system to move from $-x_\mathrm{b}$ on the lower adiabatic surface to $+x_\mathrm{b}$ on the higher adiabatic surface without returning to $-x_\mathrm{b}$ on the lower adiabatic surface.
Since each TPT takes a different value for each trajectory as a result of thermal fluctuation, the TPT has a distribution.
It is clear that the TPT in our definition depends on the value of $x_\mathrm{b}$, which is an arbitrary parameter.
To evaluate the dependence on $x_\mathrm{b}$, first, we computed TPTs with six different values of $x_\mathrm{b}$ by equally dividing the distance between the avoided crossing point and the well minimum (see Section S3 in Supplementary Information). Since the results did not vary significantly with $x_\mathrm{b}$, we choosed the value of $x_\mathrm{b}$ that is the smallest, i.e., closest to the potential crossing point.
Note that the analytical solution of the TPT distribution without considering the non-adiabatic effect ($P_{\rm{hop}} = 0$) is given by Ref.\citenum{Laleman2017}, assuming a inverted parabola energy surface$(=-\frac{1}{2}\omega'^2x^2+E_X)$ around the energy-barrier region.
We compared the calculated TPT distribution with the analytical solution given by adopting the harmonic frequency of the inverted parabola as $\omega'=\omega$.

It would be beneficial to look at the difference between the reaction time constant (MFPT) and the TPT from another perspective.
Berezhkovskii and Szabo\cite{Berezhkovskii2022} decomposed the first passage time ($t_{\rm FP}$) into a sum of the TPT ($t_{\rm TP}$) and what they call the looping time ($t_{\rm LP}$), $t_{\rm FP} = t_{\rm LP} + t_{\rm TP}$.
In equilibrium, a trajectory can depart from the product to get to the reactant. The trajectory can then depart from the reactant and re-enter the reactant region for multiple times, and finally gets back to the product region. The looping time $t_{\rm LP}$ is the difference in the first and last time of the re-entrant.
Interestingly, in the adiabatic and overdamped case, analytical results indicate that both the MFPT (and the mean looping time) and the mean TPT are expressed as functions of the free energy barrier height, but with different shapes; the MFPT increases exponentially, but the mean TPT {\it decreases} following a more slowly varying form, e.g., polynomial, with the barrier height.\cite{Kim2015,Berezhkovskii2017,Berezhkovskii2019}
Hence, the MFPT and the mean TPT differ in orders of magnitude when there is a substantial barrier. 
In such situations, the MFPT (thus the reaction rate) is governed by the waiting time in the reactant potential well for a successful reaction event,
i.e., the looping time, where each successful reaction event, the duration of which is $t_{\rm TP}$, may be considered as an incident process.
Taking this differently, with the substantial barrier, the physical meaning of the MFPT (or the reaction rate) and the mean TPT are clearly separated; the MFPT (or the reaction rate) characterizes the frequency, i.e., the probability per unit time, of successful reaction events, while the mean TPT is the measure of duration for individual reaction event, characterizing the dynamical aspect of the reaction.

\subsection{Simulation details} \label{subsec_simulationdetails}
The coordinates and velocities were propagated using the velocity Verlet algorithm.
Because the dynamics follows the Langevin equation (Eq.\eqref{7}), we adopted the revised velocity Verlet scheme, in which an velocity update includes the frictional parameter and the random force.\cite{GrnbechJensen2013}
\begin{align}
    &x_{n+1} = x_n+b\dot{x}_n\Delta t+\frac{b}{2}\ddot{x}_n\Delta t^2+\frac{b}{2}\xi_{n+1}\Delta t \notag  \\
    &\dot{x}_{n+1} = a\dot{x}_{n}+\frac{1}{2}\left(a\ddot{x}_{n}+\ddot{x}_{n+1}\right)\Delta t + b\xi_{n+1} \notag  \\
    &a = \frac{1- \frac{\gamma }{2}\Delta t}{1+ \frac{\gamma }{2}\Delta t}, \
    b = \frac{1}{1+ \frac{\gamma }{2}\Delta t}, \
    \xi_{n+1} = \int_{t_n}^{t_{n+1}} \xi(s) ds,  \notag \\
         &\left\{
    \begin{matrix}
         \langle \xi_n \rangle =  0 \\
         \langle \xi_n\xi_l \rangle = 2\gamma k_\mathrm{B}T \Delta t \delta_{n,l}
     \end{matrix}
     \right. .
\end{align}
where $t_n = n\Delta t$ denotes the discrete time, and $x_n$ represents the value of $x(t)$ evaluated at $t_n$.

To obtain statistically converged results, we used 10,000 trajectories for the population dynamics and 100,000 trajectories for the TPT analysis.
The parameters employed in the simulations are listed in Table \ref{tab:a}.
We chose the diabatic coupling $V$ and the friction parameter $\gamma$ as variable parameters. 
Conceptually, the former and latter control the non-adiabatic effects and the strength of solvent friction, respectively.
We set the remaining parameters: the reorganization energy $x_\mathrm{a}$, the temperature $T$, the harmonic frequency $\omega$, and the timestep $\Delta t$ to a fixed value according to Refs. \citenum{Landry2012} and \citenum{Goyal2015}.
Note that we also performed simulations with varied temperatures and the results indicate that the temperature dependence is insignificant (see Section S5 in Supplementary Information).
In the diabatic representation, the energy barrier of $0.10 \ \mathrm{eV}$ has to be overcome to move along the reaction coordinate from $-x_\mathrm{a}$ at the reactant state to $x_\mathrm{a}$ at the product state.
We here limit out scope to the electron transfer processes in the non-adiabatic regime,
in which the Landau--Zener parameter\cite{Nitzan2006Book}
\begin{align}
\alpha_{\mathrm{LZ}} &=\frac{2\pi|V|^2}{\hbar \dot{x}|F_2-F_1|} \label{aLZ}  \\
&\approx \frac{\pi|V|^2}{\hbar\omega^2x_\mathrm{a}\sqrt{k_\mathrm{B}T}} \label{aLZ_MB}
\end{align}
is less than 1.
Within the parameter range shown in Table \ref{tab:a}, $\alpha_{\mathrm{LZ}}$ falls within the range from $3.33\times10^{-3}$ to $8.53\times10^{-1}$.
To transform Eq.\eqref{aLZ} to Eq.\eqref{aLZ_MB} we assumed Maxwell--Boltzmann distribution of momentum at the avoided crossing point.
In addition, the values of $V$ used in this study are consistent with typical values for non-adiabatic electron-transfer reactions, supporting the physical relevance of the parameter range considered here.\cite{Blumberger2015, Umeyama2018, Lambert2024}

\begin{figure}
    \centering
    \includegraphics[width=3in]{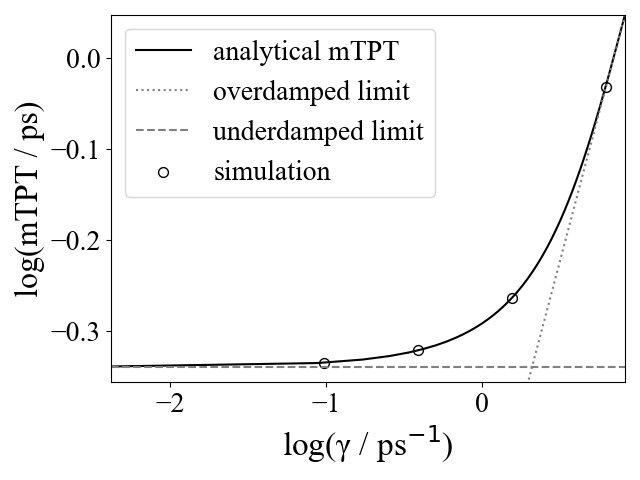}
    \caption{
    Dependence of the analytical solution of mTPT on friction parameter $\gamma$. 
    $x_\mathrm{b}=2.06 \ \mathrm{amu^{1/2}\AA}$.
    In the underdamped and overdamped limits, the behavior asymptotically approaches a straight line. 
    The dots represent the values of $\gamma$ used in the simulations.
    }
    \label{gamTPT}
\end{figure}

The parameter range of $\gamma$ was determined according to a previous study,\cite{Landry2012}
in which the values of $\gamma$ were determined to satisfy two conditions.
First, trajectories reside the avoided crossing region only for sufficiently short time compared to the waiting time around the potential well minima.
Second, the system has to be thermalized quickly after each electron transfer event.
The former and the latter conditions pose the underdamped and overdamped limit of $\gamma$, respectively.
Furthermore, the underdamped and the overdamped limit can be determined using the analytical solution for the mean TPT (mTPT).\cite{Laleman2017}
The relationship between $\gamma$ and mTPT with $x_\mathrm{b} = 2.06 \ \mathrm{amu^{1/2}\AA}$ is shown in Fig.\ref{gamTPT}, suggesting that the underdamped and overdamped limits of 
the analytical solution of mTPT falls within the range of $\gamma$ used in this study.

\begin{table}[h]
    \centering
    \caption{Parameters employed for the simulations.}
    \label{tab:a}
    \begin{tabular}{p{3cm}p{5cm}p{5cm}}
    \hline\hline
        \multicolumn{1}{c}{Parameter} & & \multicolumn{1}{c}{Range}  \\
        \hline
        \multicolumn{1}{c}{$x_\mathrm{a}$} & \multicolumn{1}{c}{ } & \multicolumn{1}{c}{$12.4 \ \mathrm{amu^{1/2}\AA}$} \\
        \multicolumn{1}{c}{$T$} & \multicolumn{1}{c}{Temperature} & \multicolumn{1}{c}{300 K} \\ 
        \multicolumn{1}{c}{$\omega$} & \multicolumn{1}{c}{Harmonic frequency} & \multicolumn{1}{c}{1.81 $\rm{ps}^{-1}$} \\
        \multicolumn{1}{c}{$V$} & \multicolumn{1}{c}{Diabatic coupling} & \multicolumn{1}{c}{$10^{-1}\sim10$ meV} \\
        \multicolumn{1}{c}{$\gamma$} & \multicolumn{1}{c}{Friction parameter} & \multicolumn{1}{c}{$10^{-2}\sim10$ $\rm{ps^{-1}}$} \\ 
        \multicolumn{1}{c}{$x_\mathrm{b}$} & \multicolumn{1}{c}{ } & \multicolumn{1}{c}{$2.06 \ \mathrm{amu^{1/2}\AA}$} \\
        \multicolumn{1}{c}{$\Delta t$} & \multicolumn{1}{c}{Timestep} & \multicolumn{1}{c}{$2.42\times10^{-2} \ \rm{fs}$} \\
        \hline\hline
    \end{tabular}
\end{table}

\section{RESULTS AND DISCUSSION}
\subsection{Population dynamics and reaction rates} \label{subsec_popandrate}

\begin{figure}
    \centering
    \includegraphics[width=3in]{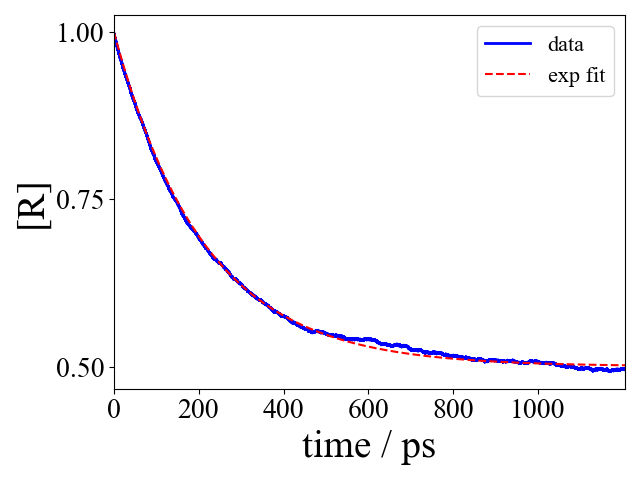}
    \caption{Time evolution of the reactant population $[\mathrm{R}]$ for $V=0.215\ \mathrm{meV}$ and $\gamma = 0.097\ \mathrm{ps^{-1}}$. The blue line represents simulation data, and the red line shows exponential fitting based on Eq.\eqref{14}.}
    \label{Rvst}
\end{figure}
 Fig.\ref{Rvst} shows the time evolution of $[\rm{R}]$. 
 The blue line represents the calculated results, and the red line is exponential fitting according to Eq.\eqref{14}.
 As expected, $[\mathrm{R}]$ approaches to $1/2$ in the long time limit.
 The same behavior was observed for all the conditions under consideration (see S2 in Supplementary Information).
\begin{figure}
    \centering
    \includegraphics[width=3in]{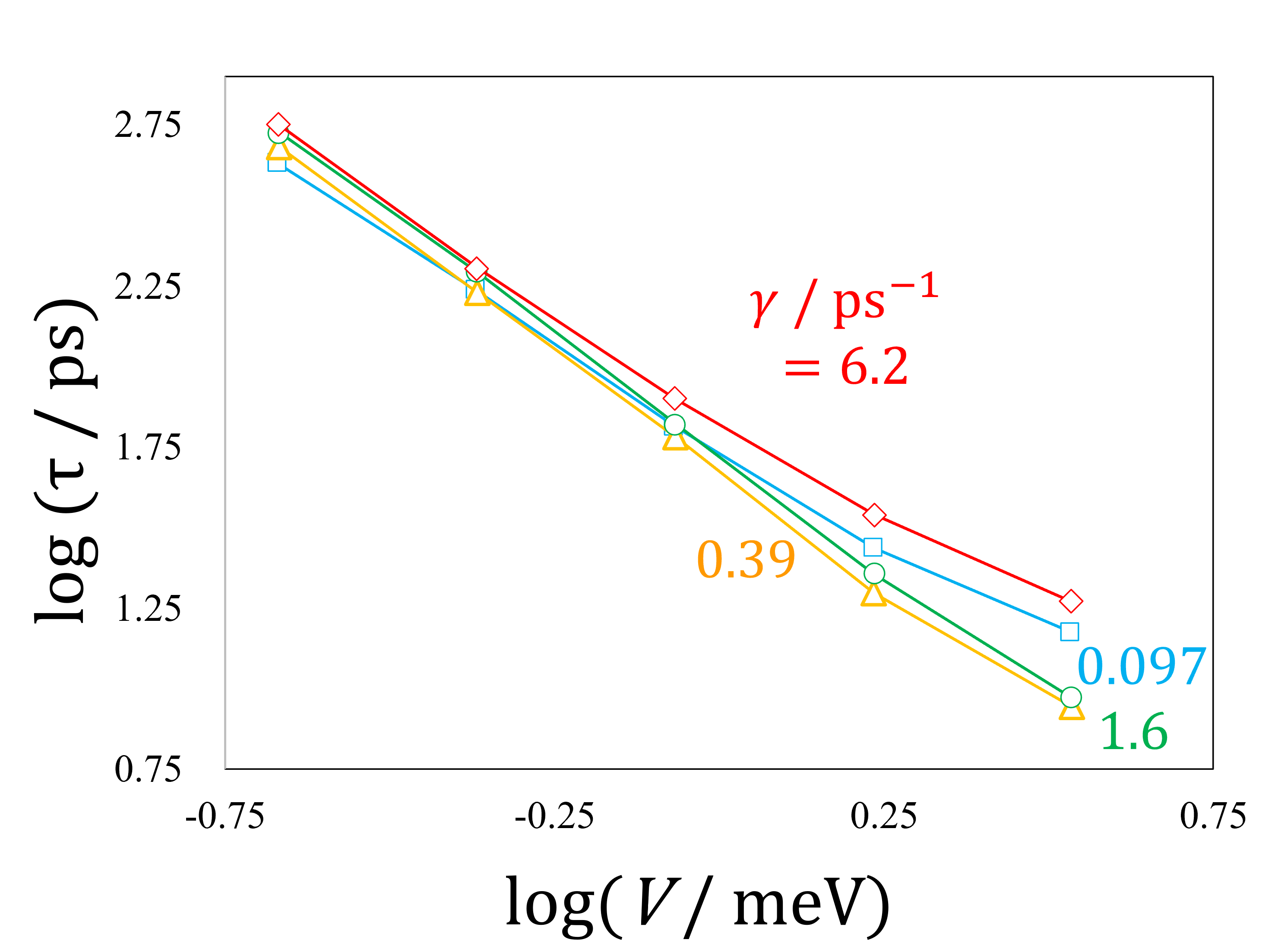}
    \caption{Dependence of time constant $\tau$ on diabatic coupling $V$. The difference in color lines corresponds the difference of value of $\gamma$.}
    \label{tauvsV}
\end{figure}

Fig.\ref{tauvsV} shows the dependence of the time constant $\tau$ on diabatic coupling $V$.
As $V$ increases, $\tau$ decreases monotonically.
This result indicates that the dependence of the hopping probability $P_{\rm{hop}}$ increases with $V$, being consistent with the results predicted by the Fermi's golden rule.
\begin{align}
    k \propto |V|^2
\end{align}
As indicated by the fact that the difference between the results at $\gamma=6.2\,{\rm ps}$ and $\gamma=0.39\,{\rm ps}$ increase with $V$, the influence of $\gamma$ becomes more substantial when $V$ is increased. Even though, the lines remain almost overlapping, suggesting that the dependence on $\gamma$ is entirely small compared with that on $V$.
In other words, in the parameter range considered here, the timescale of the overall reaction is governed by $V$.

\subsection{TPT distributions} \label{subsec_tpt}

\begin{figure}
    \centering
    \includegraphics[width=3in]{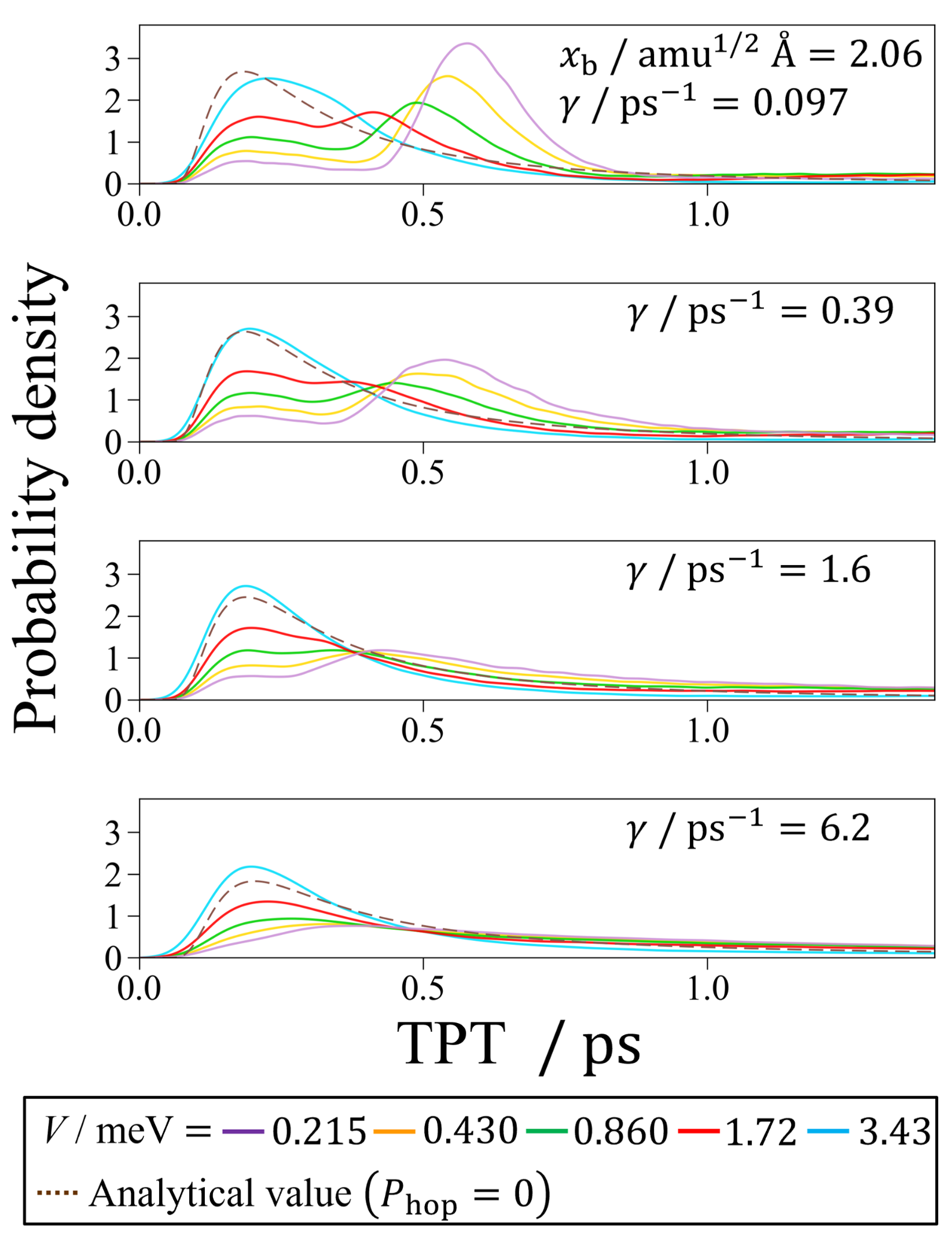}
    \caption{Distribution of transition path times (TPTs) for various values of $V$ and $\gamma$.
    The difference in line colors corresponds the difference in $V$.
    The dotted lines are the analytical values when the hopping probability $P_{\mathrm{hop}}=0$.}
    \label{TPT}
\end{figure}

\begin{figure*}
    \centering
    \includegraphics[width=5in]{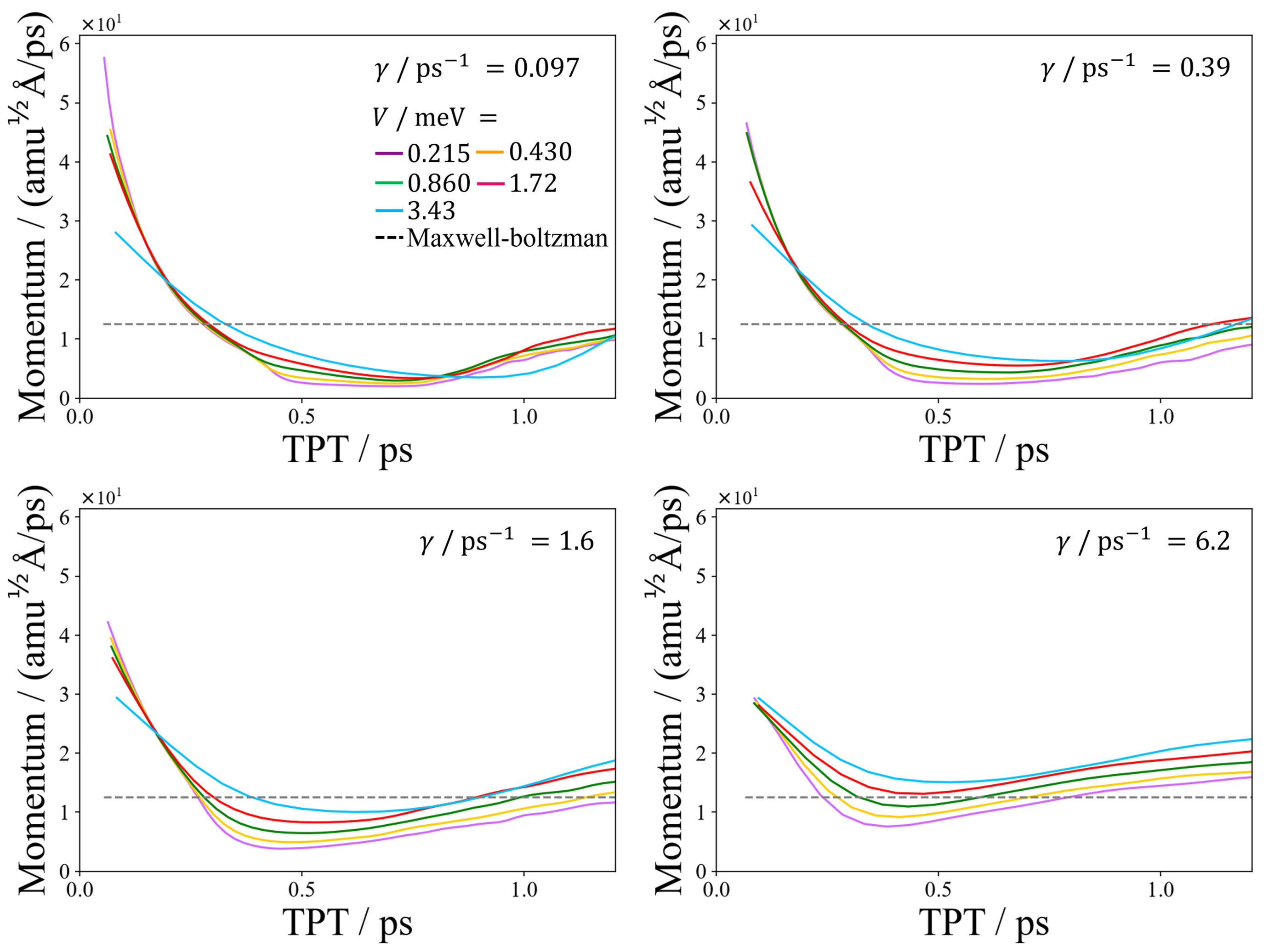}
    \caption{The mean momentum at the potential crossing point, relative to the TPT. For each trajectory, the momentum was recorded at the last event of passing through the avoided crossing point in the forward direction. 
    The difference in line colors corresponds the difference in $V$.
    The dotted line is the mean of momentum obtained from Maxwell--Boltzmann distribution.}
    \label{pTPT}
\end{figure*}

\begin{figure*}
    \centering
    \includegraphics[width=5in]{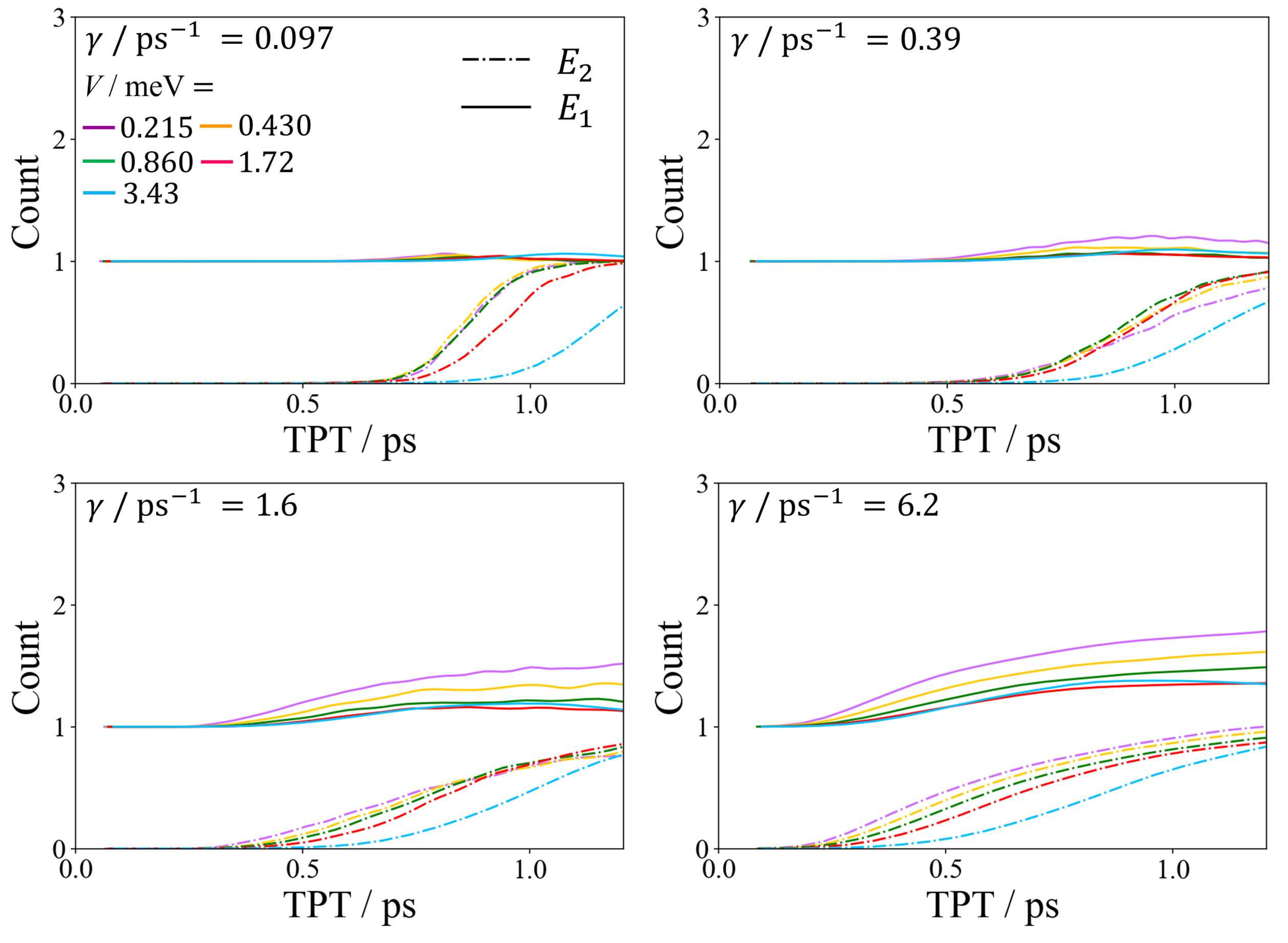}
    \caption{The mean count of passing through the potential crossing point in the forward direction is plotted relative to the TPT.
            Different line colors correspond to different values of $V$. 
            Solid and dotted lines represent the number of passages on $E_1$ and $E_2$, respectively. 
            The plot indicates the energy surface on which the trajectory was located immediately after crossing the potential crossing point.}
    \label{countTPT}
\end{figure*}

Fig.\ref{TPT} shows the dependencies of the TPT distribution on $V$ and $\gamma$, indicating that the increase in $V$ shifts the TPT distribution toward shorter times, and this trend is consistently observed in any $\gamma$.
To investigate the origin of these changes, we compared the mean momentum at the potential crossing point versus the TPT (Fig.\ref{pTPT}).
Fig.\ref{pTPT} shows that the shorter the TPT is, the larger the momentum is, and the mean momentum at the crossing point relative to TPT does not depend much on $V$.
As shown in Eq.\eqref{9}-\eqref{11}, the non-adiabatic transition probability increases, i.e., the {\it diabatic} transition probability decreases, as the momentum increases.
Therefore, when $V$ is increased, it becomes more probable for trajectories to reach $x_\mathrm{b}$ from $-x_\mathrm{b}$ while maintaining large momenta, resulting in the short-time shift of the peak of TPT distribution.

Fig.\ref{pTPT} also shows that the relation between the mean momentum at the crossing point versus the TPT is not monotonic; the mean momentum at the crossing point has decreasing trend in the short TPT region, but it turns to increase in the long TPT region.
This trend can be attributed to the increase in the number of back-and-forth movement for each TPT measurement. 
Fig.\ref{countTPT} shows the mean count of passing through the potential crossing point on the $E_1$ and $E_2$ adiabatic surfaces.
In all cases, the mean number of passages increases with increasing TPT on both $E_1$ and $E_2$.
In particular, the increase in the mean count of passages on $E_2$ suggests that non-adiabatic transitions to the upper surface tend to increase the TPT.

Moreover, increasing $\gamma$ makes the TPT distribution less dependent on $V$ (Fig.~\ref{TPT}), weakens the correlation between the TPT and the mean momentum at the potential crossing point (Fig.~\ref{pTPT}), and increases the mean count of passages through the crossing point on both surfaces (Fig.~\ref{countTPT}).
The mean momentum at the potential crossing point ultimately approaches to that obtained from the simple Maxwell--Boltzmann distribution. 
These behaviors can be understood that when the contribution of the random force $\xi$ is significant the TPT is not substantially influenced by the momenta at the potential crossing point, i.e., $\ddot{x} \approx \xi(t)$.

\begin{figure}[h]
    \centering
    \includegraphics[width=3in]{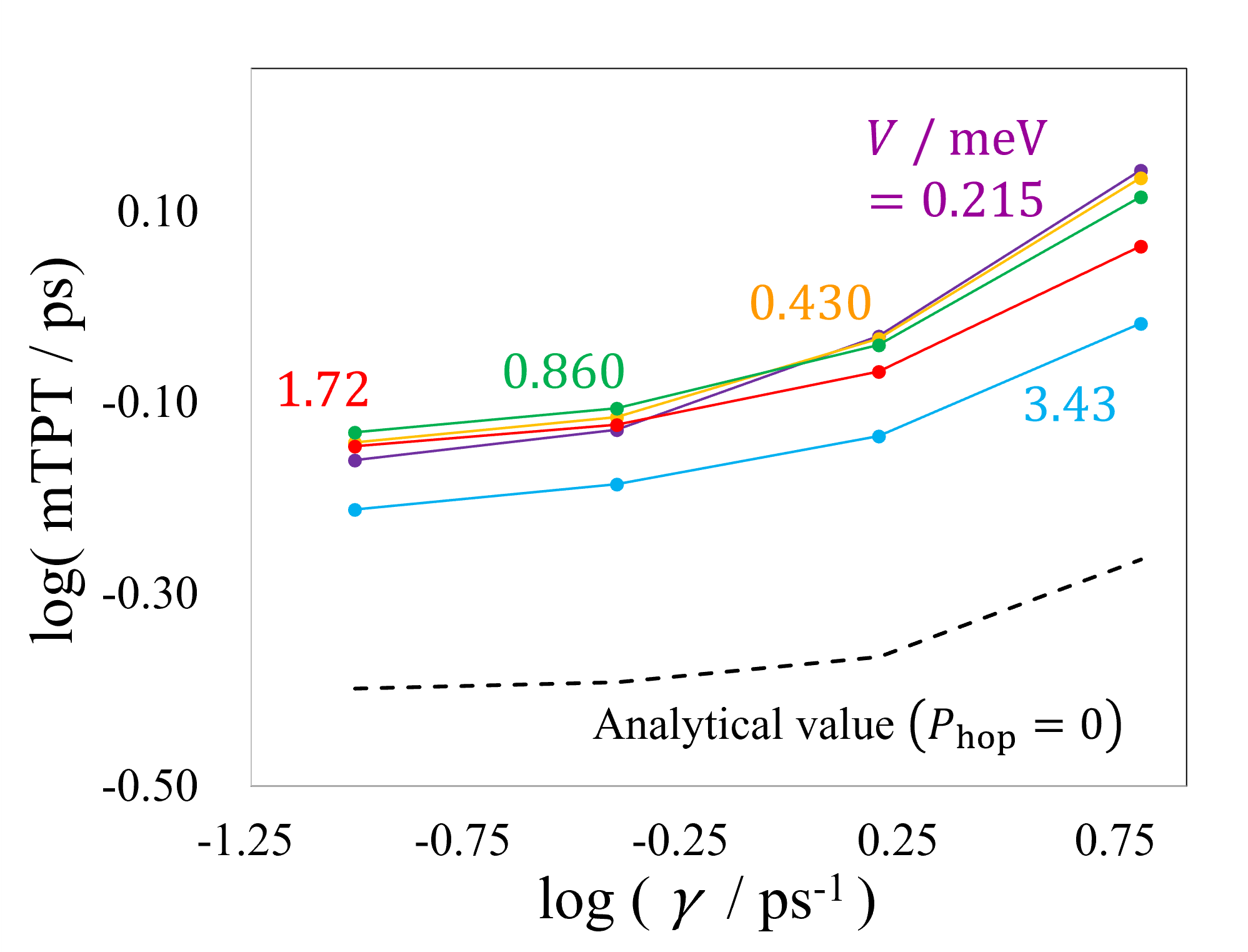}
    \caption{Dependence of $\mathrm{mTPT}$s on friction parameter $\gamma$.
    The difference in line colors corresponds the difference of value of $V$.
    The dotted line is the analytical value when the hopping probability $P_{\mathrm{hop}=0}$.}
    \label{mTPT}
\end{figure}

\subsection{Mean of TPT (mTPT)} \label{subsec_mtpt}
Fig.\ref{mTPT} shows the dependence of the mean of the $\mathrm{TPT}$ ($\mathrm{mTPT}$) on $\gamma$.
As $\gamma$ increases, the $\mathrm{mTPT}$ increases.
This means that under the diffusive regime, trajectories tend to take more time to pass through the energy-barrier region.
In addition, it can be found that the $\mathrm{mTPT}$ does not change monotonically with $V$.
The dotted line, i.e. the analytical value when the hopping probability $P_\mathrm{hop}$ is set to 0, exhibits the similar line shape with those of the simulation results suggesting that the non-adiabatic effects does not significantly change the dependence of mTPT on $\gamma$.
Based on Fig.\ref{mTPT}, $V$ has a complicated influence on the mTPTs, while the influence of $\gamma$ is monotonic, and the latter is more significant than the former. 
Thus, the $\mathrm{mTPT}$ is primarily determined by $\gamma$ in contrast to the TPT distribution.

\subsection{Comparison of mTPT and reaction time constants which depends on both $\gamma$ and $V$ substantially} \label{subsec_comparison}

\begin{figure}[h]
    \centering
    \includegraphics[width=3in]{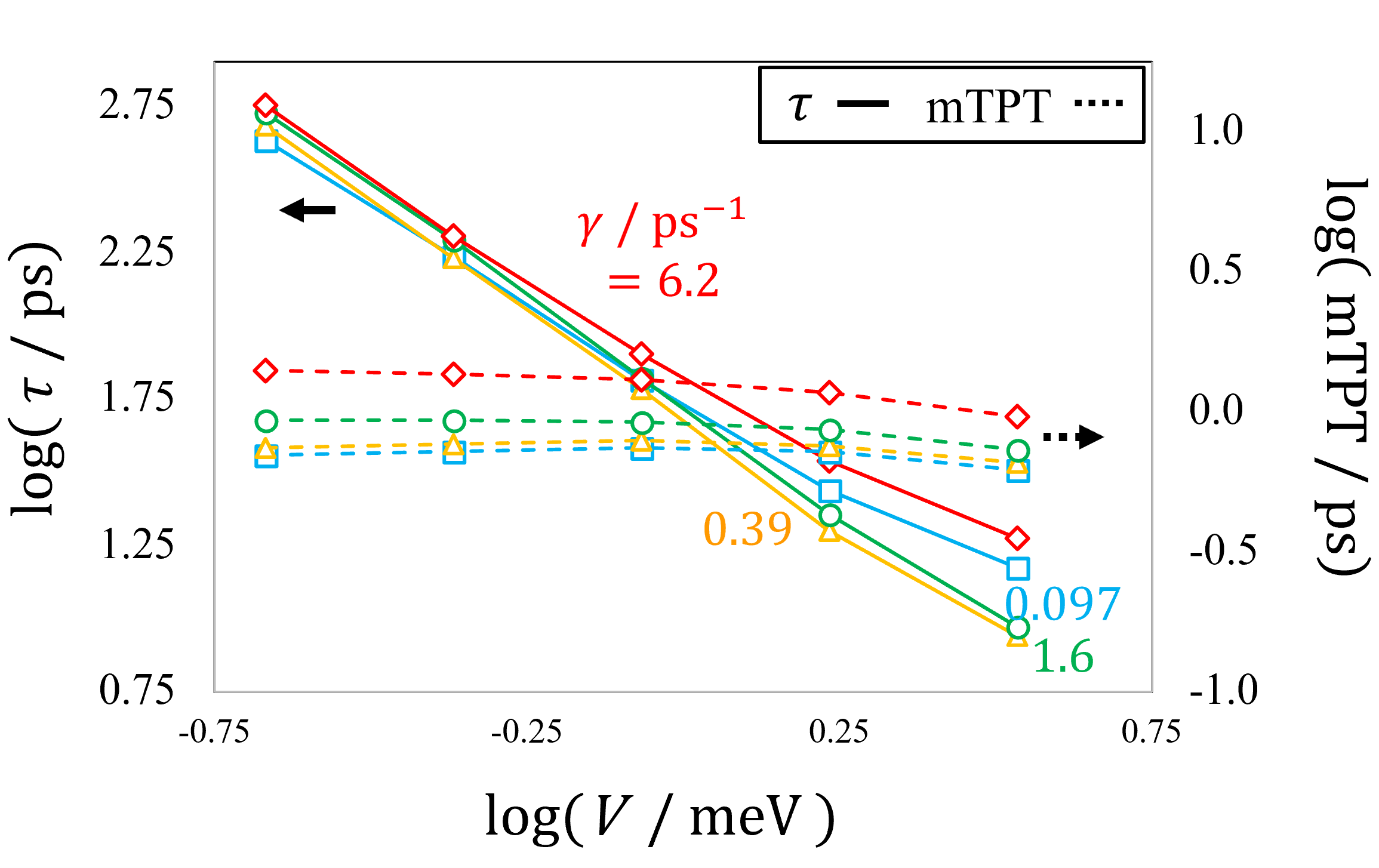}
    \caption{Dependence of $\tau$ and $\rm{mTPT}$ on $V$. The difference in line colors corresponds the difference in $\gamma$.
    The solid and dotted lines represent the $\tau$ and $\mathrm{mTPT}$, respectively.}
    \label{mTPTvstau}
\end{figure}
Fig.\ref{mTPTvstau} shows the dependence of $\tau$ and $\mathrm{mTPT}$ on $V$.
As $V$ increases, $\tau$ decreases significantly, as mentioned above, while $\mathrm{mTPT}$ remains largely unchanged.
In most cases, mTPT is smaller than $\tau$ by more than one order of magnitude.
However, for example, when $V=3.43 \ \mathrm{meV}$, the $\mathrm{mTPT}$ is greater than $5\%$ of $\tau$.
This shows that, under certain conditions, the ratio of $\mathrm{mTPT}$ to $\tau$ can be substantial even in the non-adiabatic regime.

\section{CONCLUSION}
This study elucidates the influence of non-adiabatic effects and fluctuations on the temporal structure of the transition process in condenced-phase ET reactions, by non-adiabatic dynamics simulations according to an extended Kramers model to account for non-adiabatic transitions based on Zhu-Nakamura theory.
We analyzed the timescale of microscopic transition processes (TPTs) of condensed-phase non-adiabatic ET reactions for the two-level avoided crossing model.
In particular, we investigated the effects of the diabatic coupling $V$ and the friction parameter $\gamma$ on TPTs, and further examined their relationship with the macroscopic reaction time $\tau$.
The results showed that while the TPT distribution depends on both $V$ and $\gamma$, the mean of TPT (mTPT) is mainly determined by $\gamma$. In contrast, the reaction time $\tau$ depends primarily on $V$, indicating that the timescales for microscopic transition processes and macroscopic population transfer are governed by different mechanisms.
Although the $\tau$ and the mTPT are well separated by several orders of magnitude in most cases, we found that the difference between $\tau$ and mTPT can be as close as one order of magnitude when $V$ is large.
Overall, this study offers a new perspective on understanding the microscopic time scales involved in non-adiabatic electron transfer reactions.
Moving forward, it will be necessary to analyze the actual pathways followed by these processes and to verify their generality through comparisons with more realistic molecular systems and experimental results.

\section*{SUPPLEMENTARY MATERIAL}
The supplementary material PDF file contains additional results regarding the statistical errors, population dynamics in other conditions, the dependence on $x_{\rm b}$, comparison between Zhu-Nakamura theory and FSSH, dependence on the temperature, and dependence on the driving force.

\section*{ACKNOWLEDGMENTS}
This study was supported by KAKENHI Grant Nos.~JP23H01922 and JP26H00371 from Japan Society for the Promotion of Science, and by PRESTO Grant No.~JPMJPR23Q2 from Japan Science and Technology Agency.
A part of the computational resources was provided by Research Center for Computational Science, Okazaki, Japan (26-IMS-C019).

\bibliography{reference}

\end{document}